\documentstyle[12pt]{article}

\def\baselinestretch{1.2}

\catcode`\@=11

\def\marginnote#1{}
\newcount\hour
\newcount\minute
\newtoks\amorpm
\hour=\time\divide\hour by60
\minute=\time{\multiply\hour by60 \global\advance\minute by-\hour}
\edef\standardtime{{\ifnum\hour<12 \global\amorpm={am}%
        \else\global\amorpm={pm}\advance\hour by-12 \fi
        \ifnum\hour=0 \hour=12 \fi
        \number\hour:\ifnum\minute<10 0\fi\number\minute\the\amorpm}}
\edef\militarytime{\number\hour:\ifnum\minute<10 0\fi\number\minute}

\def\draftlabel#1{{\@bsphack\if@filesw {\let\thepage\relax
   \xdef\@gtempa{\write\@auxout{\string
      \newlabel{#1}{{\@currentlabel}{\thepage}}}}}\@gtempa
   \if@nobreak \ifvmode\nobreak\fi\fi\fi\@esphack}
        \gdef\@eqnlabel{#1}}
\def\@eqnlabel{}
\def\@vacuum{}
\def\draftmarginnote#1{\marginpar{\raggedright\scriptsize\tt#1}}

\def\draft{\oddsidemargin -.5truein
        \def\@oddfoot{\sl preliminary draft \hfil
        \rm\thepage\hfil\sl\today\quad\militarytime}
        \let\@evenfoot\@oddfoot \overfullrule 3pt
        \let\label=\draftlabel
        \let\marginnote=\draftmarginnote
   \def\@eqnnum{(\theequation)\rlap{\kern\marginparsep\tt\@eqnlabel}%
\global\let\@eqnlabel\@vacuum}  }

\def\preprint{\twocolumn\sloppy\flushbottom\parindent 2em
        \leftmargini 2em\leftmarginv .5em\leftmarginvi .5em
        \oddsidemargin -.5in    \evensidemargin -.5in
        \columnsep .4in \footheight 0pt
        \textwidth 10.in        \topmargin  -.4in
        \headheight 12pt \topskip .4in
        \textheight 6.9in \footskip 0pt
        \def\@oddhead{\thepage\hfil\addtocounter{page}{1}\thepage}
        \let\@evenhead\@oddhead \def\@oddfoot{} \def\@evenfoot{} }

\def\numberbysection{\@addtoreset{equation}{section}
        \def\theequation{\thesection.\arabic{equation}}}

\def\underline#1{\relax\ifmmode\@@underline#1\else
        $\@@underline{\hbox{#1}}$\relax\fi}

\def\titlepage{\@restonecolfalse\if@twocolumn\@restonecoltrue\onecolumn
     \else \newpage \fi \thispagestyle{empty}\c@page\z@
        \def\thefootnote{\fnsymbol{footnote}} }

\def\endtitlepage{\if@restonecol\twocolumn \else \newpage \fi
        \def\thefootnote{\arabic{footnote}}
        \setcounter{footnote}{0}}  

\catcode`@=12
\relax

\def\figcap{\section*{Figure Captions\markboth
        {FIGURECAPTIONS}{FIGURECAPTIONS}}\list
        {Figure \arabic{enumi}:\hfill}{\settowidth\labelwidth{Figure
999:}
        \leftmargin\labelwidth
        \advance\leftmargin\labelsep\usecounter{enumi}}}
 \relax
\def\tablecap{\section*{Table Captions\markboth
        {TABLECAPTIONS}{TABLECAPTIONS}}\list
        {Table \arabic{enumi}:\hfill}{\settowidth\labelwidth{Table
999:}
        \leftmargin\labelwidth
        \advance\leftmargin\labelsep\usecounter{enumi}}}
 \relax
\def\reflist{\section*{References\markboth
        {REFLIST}{REFLIST}}\list
        {[\arabic{enumi}]\hfill}{\settowidth\labelwidth{[999]}
        \leftmargin\labelwidth
        \advance\leftmargin\labelsep\usecounter{enumi}}}
 \relax
\makeatletter
\newcounter{pubctr}
\def\publist{\@ifnextchar[{\@publist}{\@@publist}}
\def\@publist[#1]{\list
        {[\arabic{pubctr}]\hfill}{\settowidth\labelwidth{[999]}
        \leftmargin\labelwidth
        \advance\leftmargin\labelsep
        \@nmbrlisttrue\def\@listctr{pubctr}
        \setcounter{pubctr}{#1}\addtocounter{pubctr}{-1}}}
\def\@@publist{\list
        {[\arabic{pubctr}]\hfill}{\settowidth\labelwidth{[999]}
        \leftmargin\labelwidth
        \advance\leftmargin\labelsep
        \@nmbrlisttrue\def\@listctr{pubctr}}}
 \relax
\makeatother
\newskip\humongous \humongous=0pt plus 1000pt minus 1000pt

\newif\ifdtup

\relax

\def\be{\begin{equation}}
\def\ee{\end{equation}}
\def\ba{\begin{eqnarray}}
\def\ea{\end{eqnarray}}

\def\del{\partial}

\def\r{\rho}
\def\a{\alpha}

\def\b{\beta}

\def\g{\gamma}

\def\d{\delta}

\def\e{\epsilon}

\def\m{\mu}
\def\n{\nu}
\def\om{\omega}
\def\Om{\Omega}
\def\l{\lambda}

\def\s{\sigma}

\def\bs{\bigskip}

\def\qq{\qquad}
\def\bl{\bigl}
\def\br{\bigr}

\def\IR{\relax{\rm I\kern-.18em R}}

\def \ha {{1\over 2}}

\def \ov {\over}

\def\IR{\relax{\rm I\kern-.18em R}}

\def\IR{\relax{\rm I\kern-.18em R}}

\def\windinglike{winding--particle--like}
\def\windingmodes{winding--particle--modes}
\def\windingpart{winding--particle}

\begin{document}

\newcommand{\ber}{\begin{eqnarray}}
\newcommand{\eer}[1]{\label{#1}\end{eqnarray}}
\begin{titlepage}
\begin{center}

\hfill THU--96/29\\
\vskip -0.1cm
\hfill NIKHEF--96/013\\
\vskip -0.1cm
\hfill July 1996\\

\vskip -0.1cm
\hfill hep--th/9607203\\

\vskip .5in

{\large \bf KILLING--YANO SUPERSYMMETRY \\
IN STRING THEORY }

\vskip 0.6in

{
{\bf Frank De Jonghe, }~~
{\bf Kasper Peeters \footnote{E--mail: t16@nikhef.nl  } } }
\vskip .02in
{\em NIKHEF, Postbus 41882, 1009 DB Amsterdam, The Netherlands}

\vskip .15in

{\em and }

\vskip .15in

{\bf Konstadinos Sfetsos}
\footnote{E--mail: sfetsos@fys.ruu.nl}
\vskip .02in
{\em Institute for Theoretical Physics, Utrecht University\\
     Princetonplein 5, TA 3508, The Netherlands}

\vskip .3in

\end{center}

\vskip .6in

\begin{center} {\bf ABSTRACT } \end{center}
\begin{quotation}\noindent
The presence of Killing--Yano tensors implies the existence of
non--standard supersymmetries in point particle theories on curved
backgrounds.  In a string theoretical context these are symmetries of
the modes describing the particle--like behavior of the string.  In
the presence of isometries we show that, in addition to these, one can
also define a new type of non--standard supersymmetry among a mixture
of particle and winding modes.  The interplay with T--duality is also
examined and illustrated by explicit examples.

\vskip .5in

\end{quotation}
\end{titlepage}
\vfill
\eject

\def\baselinestretch{1.2}
\baselineskip 16 pt
\noindent


\makeatletter
\@ifundefined{explicit}{\def\details{\comment}}{\def\details{\relax}}
\@ifundefined{explicit}{\def\enddetails{\endcomment}}{\def\enddetails{\relax}}
\makeatother
\def\bref#1{(\ref{#1})}
\def\AP#1#2#3{ {{\sl Ann.\,Phys.\,}(N.Y.)\,}
    {\bf  {#1}} ({#2}) {#3}}
\def\CMP#1#2#3{ {\sl Commun.\,Math.\,Phys.\,}
    {\bf  {#1}} ({#2}) {#3}}
\def\IJMPA#1#2#3{ {\sl Int.\,J.\,Mod.\,Phys.\,}
    {\bf A{#1}} ({#2}) {#3}}
\def\JMP#1#2#3{ {\sl J.\,Math.\,Phys.\,}
    {\bf  {#1}} ({#2}) {#3}}
\def\MPLA#1#2#3{ {\sl Mod.\,Phys.\,Lett.\,}
    {\bf A{#1}} ({#2}) {#3}}
\def\NC#1#2#3{ {\sl Nuovo\,Cim.\,}
    {\bf  {#1}} ({#2}) {#3}}
\def\NPB#1#2#3{ {\sl Nucl.\,Phys.\,}
    {\bf B{#1}} ({#2}) {#3}}
\def\PL#1#2#3{ {\sl Phys.\,Lett.\,}
    {\bf  {#1}} ({#2}) {#3}}
\def\PLB#1#2#3{ {\sl Phys.\,Lett.\,}
    {\bf B{#1}} ({#2}) {#3}}
\def\PR#1#2#3{ {\sl Phys.\,Rep.\,}
    {\bf  {#1}} ({#2}) {#3}}
\def\PRL#1#2#3{ {\sl Phys.\,Rev.\,Lett.\,}
    {\bf  {#1}} ({#2}) {#3}}
\def\PRD#1#2#3{ {\sl Phys.\,Rev.\,}
    {\bf D{#1}} ({#2}) {#3}}
\def\OPR#1#2#3{ {\sl Phys.\,Rev.\,}
    {\bf  {#1}} ({#2}) {#3}}
\def\TMP#1#2#3{ {\sl Theor.\,Math.\,Phys.\,}
    {\bf  {#1}} ({#2}) {#3}}
\newcommand{\dr}{\raise.3ex\hbox{$\stackrel{\leftarrow}{\delta }$}}
\newcommand{\dl}{\raise.3ex\hbox{$\stackrel{\rightarrow}{\delta}$}}
%
\newcommand{\cA}{{\cal A}}
\newcommand{\cB}{{\cal B}}
\newcommand{\cC}{{\cal C}}
\newcommand{\cD}{{\cal D}}
\newcommand{\cE}{{\cal E}}
\newcommand{\cF}{{\cal F}}
\newcommand{\cG}{{\cal G}}
\newcommand{\cH}{{\cal H}}
\newcommand{\cI}{{\cal I}}
\newcommand{\cJ}{{\cal J}}
\newcommand{\cK}{{\cal K}}
\newcommand{\cL}{{\cal L}}
\newcommand{\cM}{{\cal M}}
\newcommand{\cN}{{\cal N}}
\newcommand{\cO}{{\cal O}}
\newcommand{\cP}{{\cal P}}
\newcommand{\cQ}{{\cal Q}}
\newcommand{\cR}{{\cal R}}
\newcommand{\cS}{{\cal S}}
\newcommand{\cT}{{\cal T}}
\newcommand{\cU}{{\cal U}}
\newcommand{\cV}{{\cal V}}
\newcommand{\cW}{{\cal W}}
\newcommand{\cX}{{\cal X}}
\newcommand{\cY}{{\cal Y}}
\newcommand{\cZ}{{\cal Z}}
%
\newcommand{\bA}{{\bf A}}
\newcommand{\bB}{{\bf B}}
\newcommand{\bC}{{\bf C}}
\newcommand{\bD}{{\bf D}}
\newcommand{\bE}{{\bf E}}
\newcommand{\bF}{{\bf F}}
\newcommand{\bG}{{\bf G}}
\newcommand{\bH}{{\bf H}}
\newcommand{\bI}{{\bf I}}
\newcommand{\bJ}{{\bf J}}
\newcommand{\bK}{{\bf K}}
\newcommand{\bL}{{\bf L}}
\newcommand{\bM}{{\bf M}}
\newcommand{\bN}{{\bf N}}
\newcommand{\bO}{{\bf O}}
\newcommand{\bP}{{\bf P}}
\newcommand{\bQ}{{\bf Q}}
\newcommand{\bR}{{\bf R}}
\newcommand{\bS}{{\bf S}}
\newcommand{\bT}{{\bf T}}
\newcommand{\bU}{{\bf U}}
\newcommand{\bV}{{\bf V}}
\newcommand{\bW}{{\bf W}}
\newcommand{\bX}{{\bf X}}
\newcommand{\bY}{{\bf Y}}
\newcommand{\bZ}{{\bf Z}}
\newcommand{\beq}{\begin{equation}}
\newcommand{\eeq}{\end{equation}}
\newcommand{\gras}[1]{\epsilon_{#1}}
\newcommand{\sdet}{\mbox{sdet}}
\newcommand{\str}{\mbox{str}}
\newcommand{\tr}{\mbox{tr}}
\newcommand{\ihbar}{\frac{i}{\hbar}}
%
\newcommand{\sqrg}{\sqrt{g}}
\newcommand{\sqrabsg}{\sqrt{\vert g \vert}}
\newcommand{\sqrabsh}{\sqrt{\vert h \vert}}
%
\newcommand{\ddr}{\raise.3ex\hbox{$\stackrel{\leftarrow}{d}$}}
\newcommand{\ddl}{\raise.3ex\hbox{$\stackrel{\rightarrow}{d}$}}
\newcommand{\wt}{\widetilde}
\def\sepand{\rule{14cm}{0pt}\and}
\def\gtwid{\raise.3ex\hbox{$>$\kern-.75em\lower1ex\hbox{$\sim$}}}
\def\ltwid{\raise.3ex\hbox{$<$\kern-.75em\lower1ex\hbox{$\sim$}}}
\newcommand{\eqn}[1]{(\ref{#1})}
%
%
\newcommand{\CD}[2]{\nabla^{(#1)}_{#2}}
\newcommand{\nit}{\noindent}
\newcommand{\vs}{\vspace{2ex}}
\def\del{\partial}
\def\marknew{\marginpar{{\sf NEW}}}
\def\markrem#1{{[\sf #1]}\marginpar{{\sf REM}}}
\def\bea{\begin{eqnarray}}
\def\eea{\end{eqnarray}}

\section{Introduction}

Extended $N=2$ and $N=4$ supersymmetry (in addition to the usual
$N=1$) in 2--dim $\s$--models has been investigated in detail
\cite{extsusy} leading to powerful structures that have applications
in many branches of modern theoretical physics from supergravity and
string theory to gravitational instantons and moduli problems in
monopole physics.

On the other hand it was pointed out in \cite{Gibbons} that in
particle models with $N=1$ world--line supersymmetry (so-called
spinning particle models), the conditions for the existence of
additional supersymmetries are different and in fact less stringent
than the corresponding ones on the world--sheet.  Roughly speaking,
they require instead of a covariantly constant complex structure to
exist on the target space only a so called Killing--Yano tensor,
satisfying $D_{\m} F_{\n\r} + D_{\n} F_{\m\r} =0$ \cite{Penrose}.
This new type of extended supersymmetry does not obey the standard
algebra and has some unusual as well as appealing features.  It can
exist in odd dimensional target spaces, as opposed to
the standard $N=2$
extended supersymmetry which requires the dimension to be even, or
to $N=4$ which restricts it to be a multiple of four.  In addition in
4--dim the signature can be Lorentzian as opposed to the standard
extended supersymmetry which only admits solutions in Euclidean or
Kleinian signature target spaces \cite{Gibbons}.  Even more intriguing
is the fact that the entire family of Kerr-Newman black holes has this
supersymmetry \cite{Gibbons}, whereas only the extremally charged
members have supersymmetry of the usual type in the context of
supergravity theory.

Since the 2--dim $\sigma$--model reduces to a particle model as far as
the center of mass motion of the string is concerned, such
Killing--Yano tensors define an approximate extended supersymmetry of
the $\sigma$--model among only the particle--like string modes.
Intuitively, under T--duality, any symmetry among the particle--like
string modes should transform into a symmetry of winding modes.
In this note we investigate the appearance of these extended
supersymmetries in the winding sector and their relation to
Target space duality (T--duality) \cite{RevT}. The latter
interpolates between effective field theories corresponding to
different backgrounds and its interplay with ordinary
supersymmetry has been fruitful in revealing string phenomena that
resolve paradoxes with field theoretical origin \cite{KoBa}.

The organisation of this paper is as follows: In section 2 we set up
our general framework in the Hamiltonian formalism, for $N=1$ as well
as extended supersymmetry, for any 2--dim $\s$--model
(technical details can be found in the appendix). Section 3
contains our main results concerning the existence of non--standard
extended supersymmetry among particle--like string modes and among
winding--particle--like modes. The behavior of these supersymmetries
under T--duality is also examined in general and by working out the
details in the 3--dim flat space and the 4--dim Taub--NUT metric and
their T--duals.
We also point out, with an example, that Killing--Yano
type supersymmetry in a string context might be important in
relation to S--duality.

\section{The $2D$ $\sigma$-model}
\label{s_2dsigma}

The action, in the component formalism,
of a 2--dim $\s$--model with $N=1$ world--sheet supersymmetry
is \cite{extsusy}:
\begin{eqnarray}
\label{susimo}
S(X,\Psi_+,\Psi_-) &=& \frac{1}{2} \int Q^+_{\mu\nu} \del_+ X^\mu
\del_- X^\nu
+ i G_{\mu\nu} \Psi_+^\mu \left(\del_- \Psi_+^\nu +
(\Omega^+)^\nu_{\lambda\rho} \del_-
X^\lambda
\Psi_+^\rho \right) \nonumber\\
& &  + i G_{\mu\nu} \Psi_-^\mu \left(\del_+ \Psi_-^\nu
+ (\Omega^-)^\nu_{\lambda\rho} \del_+ X^\lambda
\Psi_-^\rho \right)
+\frac{1}{2} R^-_{\mu\nu\rho\lambda} \Psi_+^\mu \Psi_+^\nu \Psi_-^\rho
\Psi_-^\lambda ~ ,
\end{eqnarray}
where
$G_{\mu\nu}$ and $B_{\mu\nu}$ are the metric and the antisymmetric tensor and
$Q^+_{\mu\nu}\equiv G_{\mu\nu} + B_{\mu\nu}$; for later use we also
define $Q^-_{\mu\nu}\equiv G_{\mu\nu} - B_{\mu\nu}$.
The generalized connections are:
$(\Omega^\pm)^\rho_{\mu\nu}= \Gamma^\rho_{\mu\nu} \pm {1 \over 2}
H^\rho_{\mu\nu}$, where%
\footnote{Round brackets
$(\ldots)$ denote complete symmetrisation over
all indices with total weight equal to one. Similarly, square brackets
$[\ldots]$ denote complete antisymmetrisation.}
$H_{\mu\nu\rho}\equiv  3 \del_{[\rho} B_{\mu\nu]}$ is the torsion.
The corresponding curvature tensors
are $R^\pm_{\mu\nu\rho}{}^\lambda=
2\partial_{[\nu} (\Omega^\pm)^\lambda_{\mu]\rho}
+ 2(\Omega^\mp)^{\lambda}_{\sigma[\nu}(\Omega^\pm)^\sigma_{\mu]\rho}
=R^\mp_{\rho\lambda\mu\nu}$.
Integration is over $\sigma^\pm=\ha(\s\pm \tau)$,
where $\s$ and $\tau$ are the natural spatial and time
coordinates on the world--sheet.
Since our considerations will be mainly
classical, the dilaton term has been omitted.

Our framework is based on the Hamiltonian formalism.
The  only complication in
the transition from the Lagrangian to the Hamiltonian formalism
is that we have to deal with the
second class constraints associated with the fermions.
A straightforward application of Dirac's
treatment of constraint systems yields for the Hamiltonian%
\footnote{In the rest of the paper integration
over the spatial variable
$\sigma$, wherever necessary, is understood.
We will also denote derivatives with respect to $\tau$ and $\s$ by
a dot and a prime respectively.}  \def\varpi {\Pi}
\begin{eqnarray}
\label{hamil}
H & = & {1 \over 2} G^{\mu\nu} \varpi_\mu\varpi_\nu + {1 \over 2} G_{\mu\nu}
    {X^\prime}^{\mu }{X^\prime}^{\nu}
   + {i \over 2} G_{\mu\nu} \left( \Psi^\mu_+ {\Psi^\prime}^{\nu}_+  -
  \Psi^\mu_-  {\Psi^\prime}^{\nu }_- \right) \nonumber \\
   && + {i \over 2}  (\Omega^+)_{\mu;\lambda\rho}
 \Psi^\mu_+ \Psi^\rho_+ {X^\prime}^{\lambda }
 - {i \over 2} (\Omega^-)_{\mu;\lambda\rho}
  \Psi^\mu_-  \Psi^\rho_- {X^\prime}^{\lambda }
- {1 \over 4} R^-_{\mu\nu\rho\lambda}
   \Psi^\mu_+ \Psi^\nu_+ \Psi^\rho_- \Psi^\lambda_- ~ ,
\end{eqnarray}
where $(\Om^\pm)_{\m;\n\l}\equiv G_{\m\r} (\Om^\pm)^\r_{\n\l}$
and $\Pi_\mu \equiv G_{\m\n} \dot X^\n$.
The conjugate momentum to $X^\m$ is
\be
\label{conmome}
 P_\m = \Pi_\m - B_{\m\n} {X^\prime}^\n +
{i\ov 2} \Om^+_{\n;\m\r} \Psi^\n_+ \Psi^\r_+
+ {i\ov 2} \Om^-_{\n;\m\r} \Psi^\n_- \Psi^\r_- ~ ,
\ee
and the non--vanishing basic Poisson (actually Dirac) brackets
are found to be
\ba
\label{dirbra}
&&\{X^\m(\s),P_\n(\s')\}  = \d^\m{}_{\n} \d(\s,\s')\ , \quad
\{\Psi_\pm^\m(\s),\Psi_\pm^\n(\s')\}=- i G^{\m\n} \d(\s,\s')~ ,
\nonumber \\
&& \{ P_\m(\s),\Psi_\pm^\n(\s') \}=
 \ha G_{\r\l,\m} G^{\l\n} \Psi_\pm^\r \d(\s,\s')~ ,\\
&&\{P_{\m}(\s),P_\n(\s')\} =
- {i\ov 4} G^{\r\l}  G_{\r\a,\m} G_{\l\b,\n} ( \Psi_+^\a \Psi_+^\b
+ \Psi_-^\a \Psi_-^\b ) \d(\s,\s') ~ .
\nonumber
\ea
It is worth noticing that the bracket between two momenta $P_\m$ and
$P_\n$ is non--vanishing.
In various manipulations that will follow we
found convenient to use the combinations
$\Pi_\m$ and $\Pi_{\pm\m} \equiv G_{\m\n} \del_{\pm} X^\n$
instead of $P_\m$.
Some additional Poisson brackets that proved useful are:
\ba
\label{dircomp}
&& \{\Pi_\m(\s),\Psi^\n_\pm(\s')\}=
\{\Pi_{\pm\m}(\s),\Psi^\n_\pm(\s')\}= (\Om^\pm)^\n_{\m\r} \Psi^\r_\pm
\d(\s,\s') ~  ,
\nonumber \\
&& \{  \Pi_\m(\s), \Pi_\n(\s')\}={i\ov 2}
\left(R^+_{\m\n\r\l} \Psi_+^\r \Psi_+^\l
+  R^-_{\m\n\r\l} \Psi_-^\r \Psi_-^\l
+ 2 i H_{\m\n\l} {X^\prime}^\l\right) \d(\s,\s') ~ .
\ea
The action (\ref{susimo}) is invariant under the standard $N=1$
supersymmetry transformations in the two chiral sectors
\ba
\label{trasusy}
\d X^\m & = & i \e_- \{Q_+,X^\m\} + i \e_+ \{Q_-,X^\m\}
\nonumber \\
 \d \Psi^\m_\pm & = & i \e_\mp \{Q_\pm,\Psi_\pm^\m \} +
i \e_\pm \{ Q_{\mp},\Psi^\m_\pm \} ~ ,
\ea
where $\e_\pm$ are constant anticommuting parameters and the
supercharges are given by
\beq
\label{sucha}
Q_\pm  =  - \Pi_{\pm\m} \Psi^\mu_\pm \pm {i \over 6}
H_{\mu \nu \rho} \Psi_\pm^\mu \Psi_\pm^\nu \Psi_\pm^\rho ~ .
\eeq
The corresponding supersymmetry algebra is:
\beq
\label{sualg}
  \{Q_{\pm},  Q_{\pm} \}= -2 i H_{\pm}\ ,
\quad  \{Q_+,Q_-\}=0 \ ,\quad \{Q_\pm,H_+\}= \{Q_\pm,H_-\}= 0\ ,
\eeq
where
\begin{eqnarray}
 H_{\pm} & = & {1 \over 2}  G^{\mu \nu } \varpi_\mu \varpi_\nu  +
  {1 \over 2}  G_{\mu \nu }
{X^\prime}^{\mu} {X^\prime}^{\nu} \pm  {X^\prime}^{\mu} \varpi_\mu
\pm i G_{\mu\nu} \Psi^\mu_\pm {\Psi^\prime}^{\nu}_\pm  \nonumber \\
&& \pm i (\Omega^\pm)_{\mu;\lambda\rho} \Psi^\mu_\pm \Psi^\rho_\pm
{X^\prime}^{\lambda}
- {1\over 4} R^-_{\mu\nu\rho\lambda} \Psi^\mu_+ \Psi^\nu_+ \Psi^\rho_-
\Psi^\lambda_- ~ .
\end{eqnarray}
As we have already mentioned, the Hamiltonian $H={1 \over 2} (H_+ + H_-)$ that
generates time--translations on the
world--sheet is given by (\ref{hamil}), whereas the generator of the
space--translations is ${1 \over 2} (H_+-H_-)$.

The search for new supersymmetries starts by making the Ansatz that,
in addition to (\ref{sucha}), extra supercharges of the form
\be
\label{2dextrasusy}
 Q^F_\pm = - \Pi_{\pm\m} (F^\pm)^\m{}_\n \Psi^\n_\pm
\pm {i\ov 6} C^\pm_{\m\n\r} \Psi_\pm^\m \Psi_\pm^\n \Psi_\pm^\r ~ ,
\ee
exist in each chiral sector separately. The conditions on $F^\pm_{\mu\nu}$
are well known \cite{extsusy}; the necessary algebra in our conventions can be
found
in appendix \ref{a_extra2dsusy}. One finds that the
target space tensor $(F^\pm)^\m{}_\n$
is an (almost) complex (hermitian) structure, which satisfies the
antisymmetry condition
\be
\label{comcond1}
F^\pm_{\m\n} +  F^\pm_{\n\m} = 0~  ,
\ee
is covariantly constant with respect to the generalized connection
\be
\label{comcond2}
 D^\pm_\m (F^\pm)^\r{}_\n = 0~  ,
\ee
and squares to $-{\bf 1 }$,
\be
\label{ffmin}
(F^\pm)^\m{}_\l (F^\pm)^\l{}_\n = -\d^\m{}_\n ~ .
\ee
In addition $C^\pm_{\m\n\r}$ is completely determined as%
\footnote{ We have also used that
$$
\label{estin} R^\pm_{\m\n\r\a} (F^\pm)^\a{}_\l =
R^\pm_{\m\n\l\a} (F^\pm)^\a{}_\r ~ , \qq \del_{[\m} C^\pm_{\n\r\l]} = 0~ .
$$
The first equation is the integrability condition for (\ref{comcond2})
whereas the second equation follows by explicitly writing out the
identity $d^2 F^\pm = 0$.}
\be
\label{comdd}
C^\pm_{\m\n\r}= 3 H_{\l[\m\n} (F^\pm)^\l{}_{\r]}~ .
\ee
Then the original $N=1$ world--sheet supersymmetry
is promoted to an $N=2$ one.
In cases where there exist three independent
complex structures in each sector the supersymmetry is actually an $N=4$ one.
The above requirements put severe restrictions on backgrounds that admit a
solution, as it has been extensively discussed
in \cite{extsusy}; for instance, in the absence of torsion $N=2$ ($N=4$)
extended supersymmetry requires K\"ahler (hyper--K\"ahler) manifolds.

\section{Non--standard extended supersymmetry and T--duality}
\label{s_extrasusy}

We would like to investigate the possibility of having
additional extended supersymmetries, not of the standard type,
by modifying some of the equations
\eqn{comcond1}--\eqn{ffmin}.
One possibility, is to allow
the complex structures to depend non--locally on the
target space variables. This arose naturally \cite{KoBa}
in investigations on the interplay between Abelian
\cite{bakasII,KoBa,hassand,sferesto} and non--Abelian \cite{nabsusy}
T--duality and world--sheet supersymmetry
and has its counterpart in conformal field
theory since non--local complex structures are
directly related \cite{KoBa,sferesto,nabsusy} to coset parafermions.
Such complex structures are not covariantly constant
\cite{KoBa,hassand} and equation \eqn{comcond2} is modified by a term
given in general in \cite{sferesto}.

Here we pursue an alternative line of investigation. We still insist
on the existence of local target space tensors $F^\m{}_\n$, but we relax the
condition that the new supersymmetry corresponding to it is a symmetry of
the entire string spectrum. We will essentially restrict
to the particle--like part of the spectrum as well as to the so called
\windinglike{} part.

\subsection{Supersymmetry for the particle--like modes}
\label{suPL}

Restricting to the modes that describe the particle behavior of the string
implies that we neglect all oscillatory modes for the bosonic
as well as for the fermionic degrees of freedom. Hence, we consider
the limit%
\footnote{ Without loss of generality we focus attention to one chiral
sector only. Hence, half of the fermionic degrees of freedom in \eqn{PL} and
later in \eqn{DPL} are set to zero.}
\beq
   {X^\prime}^{\mu} = 0~ ,  \qq  {\Psi^\prime}^{\mu}_- = 0 ~ ,
\qq \Psi^\mu_+ = 0 ~ .
\label{PL}
\eeq
which will call for brevity the particle limit.
We refer the reader to the appendix for the computational details and
present the results here. We find the
generalisation to nonzero torsion of the well known result \cite{Gibbons}
that the conditions for an extra supersymmetry
are \eqn{comcond1} and a modified version of \eqn{comcond2} given by
\be
\label{comcondY}
 D^-_\m F^-_{\r\n}  + D^-_\n F^-_{\r\m} = 0~  ,
\ee
(see also \cite{vHR}). The corresponding supercharge $Q^F_-$ is given by
\eqn{sucha}, with
\be
\label{CYan}
 C^-_{\m\n\r}= - 2 D^-_{[\m} F^-_{\n\r]} +
3 H_{\l[\m\n} (F^-)^\l{}_{\r]}~ ,
\ee
and it should obey
\footnote{We also have used
the integrability condition for \eqn{comcondY} which,
after using \eqn{dc}, reads
$$ \label{estaf}
D^-_\m C^-_{\a\b\g} = 3 R^-_{\m\r[\a\b} (F^-)^\r{}_{\g]}~ . $$ }
\be
 \del_{[\m} C^-_{\n\r\l]} = 0 ~ .
\label{dc}
\ee
This condition was also encountered in the case of extended
supersymmetry in the full 2--dim model (see footnote 3),
but unlike this case here it actually leads to a restriction
for the string torsion which, after using \eqn{CYan},
reads
\be
\del_{[\m} \left( H_{|\a|\n\r} (F^-)^\a{}_{\l]} \right)= 0~ .
\label{dHF}
\ee
In the case of zero torsion \eqn{comcondY}, \eqn{CYan}
reduce to the conditions for the
existence of a Yano tensor.
Explicit examples for such tensors can be found
in \cite{Gibbons,Monopole}.
Notice that condition \eqn{ffmin} is not a requirement, since the
supercharge corresponding to the new non--standard supersymmetry,
doesn't have to
square to the Hamiltonian \cite{Gibbons} (cf. \eqn{sualg}).

In our string inspired discussion, we have assumed so far that the
torsion was integrable, i.e. $H=dB$.  Since spinning particle models
are interesting in their own right it is important to know the
analogs of the conditions \eqn{dc}, \eqn{dHF} for existence of extended
Yano--type world--line supersymmetry in general models, where the
torsion is not integrable. Sparing the reader from all computational
details, it turns out that \eqn{dc} is modified to
\be
 \del_{[\m} C^-_{\n\r\l]} + \ha  F_{\a[\m\n\r} (F^-)^\a{}_{\l]} = 0
{}~ , \qq F_{\m\n\r\l} \equiv 4 \del_{[\m} H_{\n\r\l]}  ~ .
\label{dcc}
\ee
Consequently the condition on the torsion \eqn{dHF} is generalised to
\be
\del_{[\m} \left( H_{|\a|\n\r} (F^-)^\a{}_{\l]} \right)
+ \ha  F_{\a[\m\n\r} (F^-)^\a{}_{\l]} = 0 ~ .
\label{dHFF}
\ee
The above generalisation is trivial
in 4--dim target spaces. The reason is that in 4--dim the tensor
$F_{\m\n\r\l}$ has only one independent component and as a
consequence  the extra terms in \eqn{dcc}, \eqn{dHFF} (as compared to \eqn{dc},
\eqn{dHF}) are proportional to the trace of the Yano tensor $(F^-)^\m{}_\m =
0$.
Finally, it is interesting that adding abelian gauge fields
doesn't affect the above conditions at all. It restricts however
the gauge field strength similarly to what was found
in the last reference in \cite{Monopole}.

In the rest of the paper we continue our discussion with
integrable torsion.

\subsection{Supersymmetry for the \windingmodes}
\label{suDPL}

In the presence of isometries (we will, for simplicity, only consider
the case of one Abelian isometry) one expects on the basis of T--duality
that a similar non-standard extra supersymmetry should appear
in the modes dual to the particle modes.  We will work in an adapted
coordinate system, with Killing vector field $\partial/\partial X^0$,
where the background fields are $X^0$--independent. Then T--duality
boils down to a canonical transformation \cite{TCaTra} that interchanges
${X^\prime}^0 \leftrightarrow P_0$ (notice, not $\Pi_0$!).
Therefore we may try to consider
instead of the particle limit
\eqn{PL} what we will call \windingpart{} limit given by
\be
\label{DPL}
   P_0 = 0~ ,\qq  {X^\prime}^{i} = 0~ ,\qq  {\Psi^\prime}^\mu_- = 0~ ,
\qq \Psi^{\mu}_+ =0 ~ .
\ee
Notice that, because of the Killing symmetry
this is consistent with the Poisson brackets \eqn{dirbra}
as it should be.
%
%
%
%
Again referring the reader to the appendix for the computational details
we have found that the necessary conditions
for a non--standard extended supersymmetry
in the \windingmodes{} are equation \eqn{comcond1} and
\ba
D^-_\n F^-_\m{}^{(i} G^{j)\n} &=& 0 ~ ,\nonumber \\
D^-_{[k}F^-_{0]j}G^{k i} &=& 0 ~ ,\label{DKY} \\
D^-_0 F^-_{0i} &=& 0 \nonumber ~ ,
\ea
whereas the expression for $C^-_{\mu\nu\rho}$ can be obtained from
\ba
\label{cDPL}
C^-_{\m\n}{}^i & = & -2 D^-_{[\m} F^-_{\n]}{}^i
+ 3 H_{\l[\m\n} (F^-)^\l{}_{\r]} G^{\r i}
\nonumber \\
C^-_{\m\n 0} & = & -2 D^-_{[\m }F^-_{\n]0} - D_0^- F^-_{\m\n}
                 + 3 H_{\l [\m\n} (F^-)^\l{}_{0]} ~ .
\ea
Equations \eqn{DKY}, \eqn{cDPL} are the conditions for
extra non--standard
supersymmetry in the winding sector.

It should be possible to obtain the conditions \eqn{DKY}, \eqn{cDPL}
by simply dualizing \eqn{comcondY} directly.
For notational purposes we will use tilded symbols when we refer
to quantities of the dual to \eqn{susimo} $\s$--model. Let's recall the
generic way to derive the transformations of the fields, by first
noticing that Buscher's  duality rules for the metric can be cast into the
form
\beq
  \label{Tmetric}
   \tilde G_{\mu\nu} =
(A_\pm)^\alpha{}_\mu (A_\pm)^\beta{}_\nu G_{\alpha \beta} \hspace{1cm}
\eeq
(where both signs give the same result as it should be), with
\begin{equation}
\begin{array}{ccc}
{(A_\pm)^\mu}{}_\nu = \pmatrix{ \pm G_{00}^{-1} & -G_{00}^{-1} Q^\pm_{j0} \cr
                                  0           & {\delta^i}_j }~ ,
& \hspace{3mm} &
{(A^{-1}_\pm)^\mu}{}_\nu = \pmatrix{ \pm G_{00} & \pm Q^\pm_{j0} \cr
                                  0           & {\delta^i}_j } \, .
\end{array}
\end{equation}
($Q$ is defined below \eqn{susimo}).
The antisymmetric tensor transforms as
\bea
   \tilde B_{0i} &=&  \frac{G_{0i}}{G_{00}}, \nonumber \\
\label{Tanti}
   \tilde B_{ij} &=& B_{ij} + \frac{1}{G_{00}}
\left(G_{i0}B_{j0} - G_{j0} B_{i0} \right)
    \,   .
\eea
In addition the transformation of the generalized connections
are \cite{hassand}:
\beq
\label{omegatrafo}
{(\tilde \Omega^\pm)^\mu}_{\lambda\rho} = (A_{\mp})^\alpha{}_\lambda
(A_\pm)^\beta{}_\rho
(A^{-1}_\pm)^\mu{}_\nu (\Omega^\pm)^\nu{}_{\alpha\beta} +
\partial_\lambda (A_{\pm})^\alpha{}_\rho (A^{-1}_\pm)^\mu{}_\alpha ~ .
\eeq
The target space fermions transform under duality as
$ \tilde \Psi_\pm^\mu = (A^{-1}_\pm) ^\mu{}_\alpha \Psi_\pm^\alpha$.
This is deduced by demanding that the bracket between two fermions in
\eqn{dirbra}, remains invariant under the duality transformation which
for the metric acts as in \eqn{Tmetric}.
Finally, the world--sheet derivatives transform as \cite{zachos,hassand}
\beq
  \label{Tmom}
\partial_\pm \tilde X^\m =
{(A^{-1}_\pm)^\m}_\n \partial_\pm X^\n - i \partial_\r
{(A^{-1}_\pm)^\m}_\l \Psi_\pm^\r \Psi_\pm^\l ~ ,
\eeq
where the fermion bilinear arises after implementing T--duality
as a manifestly $N=1$ supersymmetric canonical transformation.
This boson--fermion symphysis is a characteristic feature of duality
in supersymmetric $\s$--models.
Imposing that the supercharge $Q^F_-$ in \eqn{2dextrasusy} has
an identical form in the T--dual model, one finds that \cite{kimro,hassand}
\be
\label{ftrafo}
\tilde F^-_{\mu\nu} =
{(A_-)^\alpha}_\mu {(A_-)^\beta}_\nu F^-_{\alpha\beta} ~ ,
\ee
and
\be
\label{ctrafo}
\tilde C^-_{\mu\nu\rho} =
{(A_-)^\alpha}_\mu {(A_-)^\beta}_\nu {(A_-)^\gamma}_\rho
C^-_{\alpha\beta\gamma}
+ 6 {(A_-)^\lambda}_{[\nu} \partial_\mu {(A^{-1}_-)^\beta}_{|\lambda}
\tilde F^-_{\beta|\rho]} ~ ,
\ee
where indices enclosed by bars are not being antisymmetrised.
Using \eqn{ftrafo}, \eqn{omegatrafo} one proves that \cite{hassand}
\be
\label{DFtrafo}
\tilde D^-_{\mu}\tilde F^-_{\nu\rho} = {(A_+)^\alpha}_\mu
{(A_-)^\beta}_\nu
{(A_-)^\gamma}_\rho D^-_{\alpha} F^-_{\beta\gamma}~ .
\ee
Hence, in the case of complex structures obeying \eqn{comcond2},
the transformed tensors are complex structures as well.
Our aim is to find the equation that the dual tensor $\tilde F^-$
satisfies, provided that the tensor $F^-$ satisfies the Killing--Yano--like
condition \eqn{comcondY}, and similarly for the tensor $\tilde C^-_{\m\n\r}$.
Using \eqn{DFtrafo} we find that
\begin{equation}
\label{ddYY}
\tilde D^-_\mu \tilde F^-_{\lambda\rho} + S^\alpha{}_\mu S^\beta{}_\lambda
\tilde D^-_\beta \tilde F^-_{\alpha\rho} =0 ~ ,
\end{equation}
where the matrix $S\equiv A^{-1}_+ A_-  = A^{-1}_- A_+$ is
defined as
\begin{equation}
S^\mu{}_\nu =  \pmatrix{  -1 &
2 B_{j0} \cr   0 & \delta^i{}_j \cr }
\end{equation}
and satisfies $S^{-1} = S$.
A more explicit form of \eqn{ddYY} is
\ba
\tilde D^-_{(i} \tilde F^-_{j)k} -
2 B_{0 (i} \tilde D^-_{j)} \tilde F^-_{0k}& =& 0~ ,\nonumber \\
 \tilde D^-_{[i} \tilde F^-_{0]j} & =&  0 ~ , \label{dualyanotrafo}
\\
\tilde D^-_0 \tilde F^-_{0i} & = & 0 ~ .\nonumber
\ea
The last two equations in \eqn{dualyanotrafo} are in one to one
correspondence with the similar equations in \eqn{DKY}
(of course written with tildes).
In order to compare the remaining ones we first have to contract
the first equation in \eqn{dualyanotrafo} by $\tilde G^{in} \tilde G^{jm}$
and then use the fact that
$B_{0i} = \tilde G_{0i}/\tilde G_{00}$ and
$\tilde G^{i0}=-\tilde G^{ij} \tilde G_{j0}/\tilde G_{00}$.
After a little algebra they become indeed identical.

The last step is to check that the transformation of $C^-_{\m\n\r}$
leads to
equations \eqn{cDPL}. In the particle limit, the Killing-Yano tensor
$F^-_{\m\n}$ and the completely antisymmetric tensor $C^-_{\m\n\r}$
are related by
(\ref{CYan}). Using this, as well as \eqn{ftrafo} and \eqn{ctrafo} one
obtains the transformed tensor:
\begin{equation}
\tilde C^-_{\mu\nu\rho} = - 2\tilde D^-_\l
 \tilde F^-_{[\nu\rho}{S^\l}_{\mu]} -
 6{S^\l}_{[\mu}{(\tilde\Omega^-)^\kappa}_{|\l|\nu}
 \tilde F^-_{|\kappa|\rho]} ~ .
\end{equation}
More explicitly
\begin{eqnarray}
\tilde C^-_{ij0} &=&
-\frac{4}{3} \tilde D^-_{[i}\tilde F^-_{j]0} + \frac{2}{3}\tilde D^-_0
      \tilde F^-_{ij} + \tilde H^\kappa{}_{ij} \tilde F^-_{\kappa 0}
   - 4 \tilde \Gamma^\kappa{}_{0[i} \tilde F^-_{j]\kappa} \,\,,
\nonumber \\
\tilde C^-_{ijm} &=& \frac{4}{3 \tilde G_{00}} \tilde D^-_0 \tilde F^-_{m[j}
                \tilde G_{i]0} - \frac{2}{3}
          \frac{\tilde G_{0m}}{\tilde G_{00}} \tilde D^-_0 \tilde F^-_{ij}
         + 3 \tilde {H^\kappa}_{[im}\tilde F^-_{j]\kappa}
         - 2 \tilde D^-_{[i}\tilde F^-_{jm]}   ~ . \label{c2}
\end{eqnarray}
Manipulations similar to those mentioned below \eqn{dualyanotrafo} together
with the observation that covariant derivatives with respect to $\tilde D_0$
contain only the
connection terms since $\tilde F^-_{\m\n}$ does not depend on $\tilde X^0$,
lead to the conclusion that $\tilde C^-_{\mu\nu\rho}$, as given by
\eqn{c2}, indeed
satisfies (\ref{cDPL}) for the dual variables, provided that conditions
\eqn{dualyanotrafo} hold.

\subsection{Examples}
\label{s_examples}

\noindent
{\em 3--dim flat space}: The simplest background where a Yano tensor exists is
3--dim flat space. In cylindrical coordinates for the metric
\beq
\label{3dm}
ds^2 = d\r^2 + \r^2 d\phi^2 + dz^2 ~ ,
\eeq
it assumes the form
\be
\label{3dY}
F =  \r^2 d\phi \wedge dz + \r z d\r \wedge d\phi~ .
\ee
Performing a duality transformation with respect to the vector field
$\del/\del \phi$,
one obtains for the dual metric and dilaton (we only mention
it for completeness) the results
\beq
\label{3ddum}
d\tilde s^2 = d\r^2 + \frac{1}{\r^2} d\tilde\phi^2 + dz^2 ~
,\qq \tilde \Phi =\ln(\r^2)
\ee
and  for the dual Yano tensor
\be
\label{3dduY}
\tilde F^- = dz \wedge d\tilde \phi
+ \frac{z}{\r} d\tilde\phi \wedge d\r~ ,
\eeq
where $\tilde F^-$ satisfies \eqn{DKY}.
The non--standard extended
supersymmetries associated with the Yano tensor \eqn{3dY} and its dual
\eqn{3dduY} are the only extended supersymmetries that can exists
for the flat 3--dim metric \eqn{3dm} and its dual \eqn{3ddum};
a usual complex structure satisfying \eqn{comcond1}--\eqn{ffmin}
requires even dimensional backgrounds.

\bs
\noindent
{\em Taub--NUT}: A more complicated example is based on
the Taub--NUT metric. The line element can be
cast into the following form
\beq
\label{TNut}
ds^2= V (d\tau + \vec \om \cdot d\vec x)^2
+ V^{-1} d\vec x^2 ~ ,
\eeq
where $\tau$ is a Killing coordinate, $\vec x=(x,y,z)$ and
\beq
V^{-1} = {1\ov 4} \bl(1+ {2m\ov |\vec x|}\br)\ ,\qq
\vec \om ={m z\ov 2 |\vec x| (x^2 + y^2)} (-y,x,0) ~ .
\eeq
The Taub--NUT metric considered as a string solution admits
$N=4$ extended supersymmetry of the usual kind
corresponding to the existence of three complex structures \cite{GIRU}.
There also exists an additional supersymmetry among the particle--like
modes. The corresponding  Yano tensor is given by
\beq
\label{YaTN}
F = 2 (d\tau + \vec \om \cdot d\vec x) \wedge
{\vec n\cdot d\vec x} + \bl(1+ {|\vec x|\ov m}\br) V^{-1}
\e_{ijk} n_i dx^j \wedge dx^k ~ , \qq \vec n = {\vec x\ov |\vec x|}~ .
\eeq
The dual to \eqn{TNut} with respect to the vector field $\del/\del\tau$
is an axionic instanton and
the explicit expression for the metric, antisymmetric tensor and
dilaton are:
\ba
\label{dTN}
d\tilde s^2 & = & V^{-1} (d\tilde\tau^2 + d \vec x^2) ~ , \nonumber \\
\tilde B & = & 2 \om_i d \tilde\tau \wedge d x^i ~ ,
\qq \tilde \Phi = \ln(V)~ .
\ea
This background has also standard $N=4$ extended
supersymmetry with complex structures given in \cite{KoBa}.
Again, as we shown in our general framework, there
exists a non--standard supersymmetry in the
\windinglike{} modes defined in \eqn{DPL} with
dual Yano tensor satisfying \eqn{DKY} and given
explicitly by
\beq
\label{duYanTN}
\tilde F^- =  V^{-1} \biggl(
- 2 d\tilde \tau \wedge \vec n \cdot d \vec x + \bl(1+
{|\vec x|\ov m}\br) \e_{ijk} n^i dx^j \wedge dx^k \biggr)~ .
\eeq
%
In the case of the Taub--NUT metric there is another Killing symmetry
distinct from the one we have discussed. It is associated with the
vector field generating rotations in the $x-y$ plane.
T--duality with respect to this vector field
breaks the manifest $N=4$ supersymmetry into
an $N=2$, with the rest being realized non--locally
\cite{KoBa}. As a general rule, if the complex
structures or the Killing--Yano's transform non--trivially
under the duality group they become non--local in the
target space variables after duality \cite{KoBa}.
In our case the Yano tensor \eqn{YaTN} is a singlet under both isometries,
thus remaining local in the dual model as well.

We have restricted our attention only to T--duality with
respect to one Killing symmetry but the extension to cases with
general Abelian or even non--Abelian group of isometries is straightforward.
Without yielding any details, it is worth mentioning that
under non--Abelian duality in the Taub--NUT metric with respect to the
$SO(3)$ isometry, the Killing--Yano supersymmetry transforms into one
for a collection of particle and
momentum modes in the dual model, similarly to
the  one we have been discussing.
The reason is that \eqn{YaTN}
is a singlet under $SO(3)$ rotations \cite{GIRU}.
This is to be contrasted with the case of ordinary
$N=4$ extended supersymmetry which under the same $SO(3)$--duality
transformation
breaks down to $N=1$ with the rest being
realized non--locally \cite{nabsusy}.

One would also like to know the behavior of Killing--Yano supersymmetry
under S--duality. We have not investigated this
question in full generality so that we will resort to an example
to point out important differences with the cases of
usual extended supersymmetry.
Let us recall that ordinary
extended supersymmetry may be destroyed under S--duality as shown with
examples in \cite{bakasII} and its fate is not clear as is with
T--duality where, as already mentioned, it just becomes non--locally
realized.
Let us consider 4--dim flat space with metric given by
\eqn{3dm} plus the term $d\tau^2$ corresponding to the fourth coordinate.
A series of T--S--T transformations, where T--duality is
performed with
respect to the rotational Killing vector field $\del/\del \phi$
gives again a vacuum
solution to Einstein's equations in accordance with the general statement in
\cite{bakasII} where the equivalence of this sequence of dualities
and the Geroch transformation
on Einstein vacuum solutions with one Killing symmetry was also shown.
The metric we found is
\be
\label{dsTST}
ds_{TST}^2 = {\r^2 \ov 1-C^2 \r^4} (d\phi- 4 C z d\tau )^2 +
(1-C^2  \r^4) (d\tau^2 + dz^2 + d\r^2)~ ,
\ee
where $C$ is a constant parameterizing the non--trivial
$SL(2,\IR)$ group element of the Geroch transformation.
The original flat background admits $N=4$ extended world--sheet supersymmetry.
It can be readily checked that this is not the case
for the background corresponding to \eqn{dsTST} which
does not have ordinary extended supersymmetry at all.
Nevertheless, a straightforward computation
reveals that it has a Yano supersymmetry with
\be
F_{TST} = C (1-C^2 \r^4) \r^2 d\tau \wedge dz - 4 C \r z d\tau \wedge d\r + \r
d\phi \wedge d\r ~ .
\ee
The same behavior is expected in more general backgrounds in the vicinity
of fixed points where the metric can be locally approximated by the flat one
written in polar coordinates.
The above example suggests that non--standard, Yano--type supersymmetries
might be important when discussing S--duality in a string context.

\section{Conclusions}

We have discussed the conditions necessary for the existence of
approximate symmetries in 2--dim $\s$--models between collections of
string modes.  We considered in detail extended supersymmetry among
the particle--like modes of the string center of mass as well as among
collections of winding and particle--like modes, where we found a new
type of supersymmetry. These supersymmetries, though completely
different from a field theoretical point of view, are naturally
unified under the action of T--duality.
Generalization of our discussion by including gauge or other
background fields seems conceptually straightforward.

It would be interesting to explore further the behavior of Yano--type
supersymmetry under S-- or T--S--T--duality and investigate the
possibility that it appears as a remnant of ordinary extended
supersymmetry after the latter is being destroyed in a conventional
sense. This could be relevant in confirming whether or not $S$--duality
is really a symmetry in string theory.

\vs\vs
\nit\hbox{{\bf Acknowledgments}\hfill}
\vs

\nit
We thank J.W. van Holten for discussions on spinning particle
models and Killing--Yano tensors.
The research of F.D.J. is supported by the Human Capital and Mobility
Program through the network on {\em Constrained Dynamical Systems}\/.
The research of K.P. is supported by Stichting FOM.
The work of K.S. was carried out with the financial support
of the European Union Research Program
{\em Training and Mobility of Researchers} and is part of the project
{\em String Duality Symmetries and Theories with
Space--Time Supersymmetry}, under contract ERBFMBICT950362.

\appendix
\section{Extra susy for the 2d model}
\label{a_extra2dsusy}

In this appendix we provide some of the technical details relevant to
the derivation of the conditions for the existence of an extended
supersymmetry that were omitted in the main text.

Conservation of the extra supercharge \eqn{2dextrasusy} implies
a vanishing Poisson bracket with the Hamiltonian \eqn{hamil}.  A
straightforward but tedious computation gives
\begin{eqnarray}
0 &=& \{ Q^-_F , H \} \nonumber \\
\label{qf-h}
  &=&  2 D^-_{[\mu} F^-_{\lambda]\alpha} \varpi^\mu {X'}^\lambda
  \Psi^\alpha_-
    - D^-_\alpha F^-_{\beta\nu} \Psi_-^\nu \left( \varpi^\alpha\varpi^\beta -
    {X'}^\alpha
          {X'}^\beta\right) \nonumber\\
  & & + \frac{i}{2}{(F^-)^\alpha}_\beta \Psi^\beta_- {X'}^\lambda
\left( R^+_{\a\lambda\mu\rho}\Psi_+^\mu \Psi_+^\rho
 - R^-_{\a\lambda\mu\rho}\Psi_-^\mu \Psi_-^\rho \right)
           \nonumber\\
  && +\frac{i}{2} ({X'}^\alpha - \varpi^\alpha ) (F^-)^\r{}_{\alpha}
  R^-_{\mu\nu\l\rho}\Psi_+^\mu \Psi_+^\nu \Psi^\lambda_-
   \nonumber \\
& & - \frac{i}{2}{(F^-)^\alpha}_\beta \Psi^\beta_-  \Pi^\l
\left( R^+_{\a\lambda\mu\rho}\Psi_+^\mu \Psi_+^\rho
 + R^-_{\a\lambda\mu\rho}\Psi_-^\mu \Psi_-^\rho \right)
           \nonumber\\
  & &  - {1 \over 4} \left( D^+_\alpha R^-_{\mu\nu\rho\lambda} +
 2{H^\sigma}_{\alpha\rho}R^-_{\mu\nu\s\l}\right)
  {(F^-)^\alpha}_\beta
  \Psi^\beta_-
     \Psi_-^\rho \Psi_-^\lambda \Psi_+^\mu \Psi_+^\nu \nonumber \\
  &&    + \frac{1}{4}C^-_{\mu\nu\rho}
          R^+_{\alpha\beta\gamma\delta}G^{\mu\alpha}\Psi_-^\nu \Psi_-^\rho
          \Psi_-^\beta
          \Psi_+^\gamma\Psi_+^\delta \nonumber\\
  & & -\frac{i}{6} \left( \varpi^\alpha + {X'}^\alpha \right)
 D^-_\alpha C^-_{\mu\nu\rho}\Psi_-^\mu \Psi_-^\nu \Psi_-^\rho ~ ,
\end{eqnarray}
where $\varpi^\mu \equiv \dot X^\m $.
Independence of the new supersymmetry from the original $N=1$ implies
the stronger conditions $\{ Q^-_F , Q^\pm \} = 0$. Notice that
the supersymmetry algebra
(\ref{sualg}) implies conservation of $Q^-_F$ from these stronger conditions.
The Poisson brackets of the candidate extra supercharge $Q^-_F$ with the two
standard supercharges are slightly easier to compute. One obtains
\begin{eqnarray}
\label{qf-q+}
\{ Q^-_F, Q^+ \} &=&
   \left( {X'}^\alpha -\varpi^\alpha \right) D^-_\mu F^-_{\alpha \beta}
   \Psi_+^\mu \Psi_-^\beta
       \nonumber \\
   && - \frac{i}{2} \left( R^+_{\alpha\mu\rho\tau} \Psi_+^\rho \Psi_+^\tau
                     + R^-_{\alpha\mu\rho\tau} \Psi_-^\rho \Psi_-^\tau
                     \right)
                 \Psi_+^\mu (F^-)^\alpha {}_\beta \Psi_-^\beta \nonumber \\
   && + {i \over 6} D^-_{\mu} C^-_{\alpha\beta\gamma} \Psi_-^\alpha
\Psi_-^\beta
    \Psi_-^\gamma \Psi_+^\mu \\
   && - {i \over 6} D^+_{\alpha} H_{\mu\nu\rho} \Psi_+^\mu
\Psi_+^\nu  \Psi_+^\rho
     (F^-)^\alpha {}_\beta \Psi_-^\beta \nonumber
\end{eqnarray}
and
\begin{eqnarray}
\label{qf-q-}
&&\{ Q^-_F, Q^- \} =
   - i \left( \varpi^\alpha \varpi^\beta + {X'}^\alpha {X'}^\beta
              - {X'}^\alpha\varpi^\beta - \varpi^\alpha {X'}^\beta \right)
              F^-_{\alpha\beta}
    + (  F^-_{\alpha\beta} + F^-_{\beta\alpha} ) {\Psi'}^\alpha_- \Psi_-^\beta
    \nonumber \\
   && \qq\qq
+ \frac{i}{2} \left( R^+_{\m\alpha\rho\tau} \Psi_+^\rho \Psi_+^\tau
                     + R^-_{\m\alpha\rho\tau} \Psi_-^\rho \Psi_-^\tau \right)
                 \Psi_-^\mu (F^-)^\alpha {}_\beta \Psi_-^\beta \nonumber \\
  && \qq\qq
+ \varpi^\alpha \Psi_-^\mu \Psi_-^\beta \left[ D^-_\beta F^-_{\alpha\mu}
   - H^\rho{}_{\beta \alpha} F^-_{\rho\mu} + {1\over 2} F^-_{\alpha\rho}
   H^\rho{}_{\mu \beta}
   + {1\over 2} C^-_{\alpha \mu \beta} \right] \\
 && \qq\qq
- {X'}^\lambda \Psi_-^\mu \Psi_-^\beta \left[  D^-_\beta F^-_{\lambda \mu} +
   D^-_\lambda F^-_{\mu \beta} +
 (\Omega^-)^\alpha{}_{\lambda\beta} F^-_{\mu\alpha}
   + (\Omega^-)^\alpha{}_{\beta \lambda} F^-_{\alpha\mu} + {1\over 2}
   F^-_{\lambda \alpha} H^{\alpha}{}_{\mu \beta}
+ {1\over 2} C^-_{\lambda\mu \beta}  \right] \nonumber \\
 && \qq\qq
+ \left[ {i \over 6} (F^-)^\rho{}_\mu D^-_\rho H_{\alpha\beta\gamma}
   + {i \over 6} D^-_\mu C^-_{\alpha\beta\gamma} + {i \over 4}
   C^-_{\rho\alpha\beta} H^\r{}_{\gamma\mu} \right]
\Psi_-^\alpha \Psi_-^\beta
    \Psi_-^\gamma \Psi_-^\mu \nonumber \, .
\end{eqnarray}
Notice that this last bracket provides a definition of
$C^-_{\m\n\r}$ in terms of $F^-_{\m\n}$
when $\{ Q^-_F, Q^- \}=0$ is imposed.

Let us briefly show how to recover the conditions for
extended supersymmetries from these expressions.  The vanishing of the
first line in (\ref{qf-h}), linear in the fermions, leads to the
condition \eqn{comcond2} that the tensor $F^-_{\mu\nu}$ is covariantly
constant. The vanishing of the first line in (\ref{qf-q-}) forces
$F^-_{\mu\nu}$ to be {\it antisymmetric}, both for the full 2d
$\sigma$-model, as well as for the limiting cases discussed
below. Finally, the vanishing of the terms quadratic in the fermions
in (\ref{qf-q-}) provides the relation between the antisymmetric
$C^-_{\mu\nu\rho}$-tensor and $F^-_{\mu\nu}$, given in \eqn{comdd}.
The vanishing of terms with more than two fermions follows from the
integrability of the previously
mentioned conditions and the Bianchi identity
$D^\pm_{[\m} R^\pm_{\n\r]\l\a} = \pm H^\b{}_{[\m\n} R^\pm_{\r]\b\l\a}$.

In the particle--limit (PL) defined by \eqn{PL}, the brackets reduce to
\bea
\label{qf-hPL}
\left.\{ Q^-_F , H \}\right|_{PL} &=&
- D^-_\alpha F^-_{\beta\nu} \Psi_-^\nu  \Pi^\alpha\Pi^\beta
    -\frac{i}{2} {(F^-)^\alpha}_\beta \Psi^\beta_- \Pi^\l
     R^-_{\alpha\l\m\r} \Psi_-^\m \Psi_-^\r
     -\frac{i}{6}\Pi^\l D^-_\l C^-_{\mu\nu\rho}
     \Psi_-^\mu \Psi_-^\nu \Psi_-^\rho ,\nonumber \\
\label{qf-q-PL}
\left.\{ Q^-_F , Q^- \}\right|_{PL} &=&
 \Pi^\alpha \Psi_-^\mu \Psi_-^\beta \left[ D^-_\beta F^-_{\alpha\mu}
   - H^\rho{}_{\beta \alpha} F^-_{\rho\mu} + {1\over 2} F^-_{\alpha\rho}
   H^\rho{}_{\mu \beta}
   + {1\over 2} C^-_{\alpha \mu \beta} \right] +
 \mbox{4-fermion term} ~ ,
\label{4F}
\eea
where the 4--fermion term is made up by the relevant terms in the
second and fifth lines in \eqn{qf-q-}. From these,
one derives conditions \eqn{comcondY} and \eqn{CYan} immediately.
Footnote 5 can be read off from the 3--fermion terms in
the first equation in \eqn{qf-hPL}. Explicit
computation using \eqn{CYan} and successive application of the Yano equation
leads to
\bea
D^-_\m C^-_{\a\b\g} &=& 3 R^-_{\m\r[\a\b} (F^-)^\r{}_{\g]} +
  2D^-_\s F^-_{\a[\g}H^\s{}_{\m]\b} -2D^-_\s
F^-_{\b[\g}H^\s{}_{\m]\a}\nonumber\\
&&+\left( 3 R^-_{[\a\b\m]}{}^\s F_{\s\g} - (\gamma\leftrightarrow\mu)\right)
  +\left( 3R^-_{[\m\g\a]}{}^\s F_{\s\b} -
(\alpha\leftrightarrow\beta)\right)~ .
\eea
The extra terms (compared to footnote 5) can be shown to be
proportional to \eqn{dHF}, and they vanish by virtue of the condition
\eqn{dc} which is obtained from the 4--fermion terms in \eqn{qf-q-PL}.
In the various manipulations the use of identities such as $D^-_\m H_{\n\l\r}=
3 R^-_{\m[\n\l\r]}$ and $D^-_{[\m} H_{\n\r]\l}= R^-_{[\m\n\r]\l} -
H_{\a\l[\m} H_{\n\r]}{}^\a$ is necessary.

Similarly the expressions in the \windingpart{} limit (WPL)
 defined by \eqn{DPL} are obtained by using $P_0=0$ to write
\beq
\label{momDPL}
\left. \Pi^\alpha\right|_{WPL} =
G^{\alpha \beta} \left.\Pi_{\beta}\right|_{WPL} =
G^{\alpha i} P_i + G^{\alpha i} B_{i 0} {X'}^0 -
 {i \over 2} G^{\alpha \mu} (\Omega^-)_{\beta ; \mu \rho}
\Psi^\beta_- \Psi^\rho_- ~  .
\eeq
Inserting this in the brackets leads to the conditions \eqn{DKY},
from respectively the $P_iP_j$ term,
the $P_i {X'}^0$ term and the ${X'}^0 {X'}^0$ term
in the $\left.\{Q^-_F, H\}\right|_{WPL}$ bracket. The bracket
$\left.\{Q^-_F, Q^-\}\right|_{WPL}$ gives again the expression for
$C^-_{\mu\nu\rho}$, as shown in \eqn{cDPL}.


\end{document}